\documentstyle[aps,multicol,prl]{revtex}

\draft

\begin{document}

\newcommand{\be}{\begin{eqnarray}}
\newcommand{\ee}{\end{eqnarray}}

\title{Fractional Calculus and the Evolution of Fractal Phenomena}

\author{Andrea Rocco and Bruce J. West}

\address{Center for Nonlinear Science, University of North Texas, P.O. Box 5368, Denton, Texas 76203-5368}

\date{\today}

\maketitle

\begin{abstract}
It is argued that the evolution of complex phenomena ought to be described
by fractional, differential, stochastic equations whose solutions have
scaling properties and are therefore random, fractal functions. To support
this argument we demonstrate that the fractional derivative (integral) of a
generalized Weierstrass function (GWF) is another fractal function with a
greater (lesser) fractal dimension. We also determine that the GWF is a
solution to such a fractional differential stochastic equation of motion.
\end{abstract}


\begin{multicols}{2}

\section{Introduction}

Significant changes in our knowledge of how to analyze nonlinear static and
dynamical phenomena in the physical and biological sciences have occurred
over the past thirty years\cite{mandelbrot,West3,meakin,schroeder}. In the
physical sciences methodology has moved away from complete reliance on the
tools of linear, analytic, quantitative mathematical physics towards a
combination of nonlinear, numerical and qualitative techniques\cite{meakin}.
Not only have many of the linear analytical approaches often proved to be
inadequate, but the entrenched geometry of Euclid, classically used to
describe natural phenomena, has not always been adequate to the task. In the
1960s Mandelbrot began discussing a new geometry of nature, one that
embraces the irregular shapes of objects such as coastlines, lighting bolts,
clouds and molecular trajectories. A common feature of these objects, which
Mandelbrot called fractals, is that their boundaries are so irregular that
it is not easy to understand how to apply simple metrical ideas and
operations to them\cite{mandelbrot,meakin}. Towards this end we shall
consider some of the metric peculiarities of a few unusual mathematical
objects, fractal functions, and discuss the possible physical implications
of their evolution in time in terms of the phenomena they are used to model.

One of the defining properties of a fractal function is that it does not
possess a characteristic scale length and consequently its derivatives
diverge. The Weierstrass function was the first example of a function that
is continuous everywhere, but is nowhere differentiable. In 1926 Richardson
\cite{richardson} suggested that this function, or one sharing its
non-analytic properties, be used to describe the velocity field of the
atmosphere, because of the apparent impossibility of modeling the irregular
flow of the wind using differential equations. It was specifically this
property of non-differentiability that Richardson believed captured the
essential features of turbulence. This view of turbulence, rather than
becoming quaint and out of date, has been demonstrated to be quite modern
\cite{shlesinger} and in fact anticipated the introduction of fractals into
the description of complex phenomena \cite{mandelbrot,meakin}.

During the same period L\'{e}vy\cite{levy} was working to establish the most
general properties necessary for a process to violate the then accepted form
of the Central Limit Theorem and still converge to a limit distribution. He
was quite successful, establishing the class of infinitely divisible
distributions, which as its name implies concerns processes whose
statistical properties persist at each level of aggregation of the data, and
are today called $\alpha $-stable L\'{e}vy processes. A deep connection
between L\'{e}vy stable processes and the Weierstrass function was
established using random walk concepts\cite{hughes,montroll1,montroll2} and
subsequently used to understand turbulent fluid flow\cite{shlesinger}.

It is not only macroscopic phenomena such as turbulence that suffer from the
problem of not being describable by analytic functions, however. The theory
of Brownian motion as formulated in 1908 by Langevin\cite{langevin} has the
form

\begin{equation}
\frac{du\left( t\right) }{dt}=-\lambda u\left( t\right) +F\left( t\right) 
\label{langevin}
\end{equation}
where $u\left( t\right) $ is the velocity of the Brownian particle, $\lambda$ 
is the dissipation rate and $F\left( t\right) $ is the random influence of
the bath of lighter particles (molecular impacts) on the Brownian particle.
In 1909 Perrin\cite{perrin} observed that the path of a Brownian particle,
as seen through his microscope, is continuous but not differentiable.
Therefore such a path can not be described by an analytic function. In 1942
Doob\cite{doob} proved that the velocity of the Brownian particle is
discontinuous for the Ornstein-Uhlenbeck process\cite{orn} described by 
(\ref{langevin}) and discusses how one constructs a stochastic differential
equation to replace the more familiar differential equations of physics up
to that time. In particular he replaced (\ref{langevin}) with 
\begin{equation}
du\left( t\right) =-\lambda u\left( t\right) dt+dB\left( t\right) 
\label{stoch}
\end{equation}
where $dB\left( t\right) $ is the increment of a Wiener process. This new
mathematics has become so accepted that we now write the Langevin equation 
(\ref{langevin}), but interpret it as the Doob equation (\ref{stoch}). A
process appears random on the macroscopic level due to the separation of
time scales in the microscopic and macroscopic processes. When such a
separation exists the Langevin equation, using the interpretation of Doob,
adequately describes the dynamics of the physical phenomenon. On the other
hand, when this separation of time scales does not exist, ordinary
statistical physics is no longer adequate to describe the phenomenon, as
discussed by Grigolini \textit{et al.}\cite{grigolini}. In particular, a lack
of time-scale separation induces a fractional, stochastic, differential
equation on the macroscopic level\cite{West3}.

What the Weierstrass and other non-analytic functions and the L\'{e}vy
distribution have in common is the lack of a differential description of the
time evolution of the phenomena which they are intended to model. It was
only a decade ago that it was recognized by Shlesinger \textit{et al}.
\cite{shlesinger} that L\'{e}vy flights can be used to describe the velocity
field of turbulent flow. The key to this understanding is that the equation
of evolution of the probability density is a fractional diffusion equation
whose solution is a truncated L\'{e}vy distribution\cite{West1}.

Fractional diffusion equations have been used to model the evolution of
stochastic phenomena with long-time memory, that is, phenomena with
correlations that decay as inverse power laws rather than exponentially in
time\cite{West3,West1,compte,chaves}. It has been known for quite some time
in the economics literature that such statistical properties can be
successfully described using finite difference equations in which the finite
difference is of fractional order, see Hosking\cite{hosking,bernan}. The
continuum limit of such fractional difference stochastic equations are
fractional differential stochastic equations,

\begin{equation}
\frac{d^{\beta }u\left( t\right) }{dt^{\beta }}=F\left( t\right)
\label{fracdiv}
\end{equation}
where here the mathematics has been generalized to the fractional calculus
to accommodate the non-analytic properties of the underlying phenomena. The
evolution of the probability density associated with the velocity in 
(\ref{fracdiv}) is a fractional partial differential equation in the phase 
space for the phenomenon, see for example West and Grigolini\cite{West2}.

Herein we suggest that the dynamics of complex phenomena, described by
fractal functions, can be expressed in terms of fractional differential
equations of motion. One approach to doing this has been adopted by
Nonnenmacher and colleagues\cite{Nonn}, who generalize the traditional
models of viscoelastic phenomena using fractional initial value problems.
The solutions to their fractional equations of motion yield physical
observables, such as the stress relaxation, that is in excellent agreement
with experiment. Others have generalized the analysis of wave phenomena,
starting with the diffractals of Berry\cite{Berry}, up to and including the
fractional wave equation of Schneider and Wyss\cite{Wyss}.

Our goal here is more modest in that we do not attempt to describe any
particular phenomenon in complete detail. On the other hand it is more
ambitious in that we demonstrate the generic result that fractal functions
have fractional derivatives and therefore complex phenomena having a fractal
dimension are more reasonably modeled using fractional equations of motion
than they are using ordinary differential equations of motion such as given
by Newton's laws.

Even sharing the universally accepted view according to which Newton's laws
apply to the fundamental constituents of complex phenomena\cite{footnote2},
this microscopic level of description is often not the natural one for the
phenomena being studied. For example, one rarely uses particle dynamics to
describe the viscous fluid flow in turbulence, instead one more often uses
the Navier-Stokes equation to describe the behavior of the velocity field
\cite{monin}. The familiar, ordinary differential equations of motion for
individual particles need to be replaced with fractional differential and
integral equations for the appropriate field variables. We show, using a
generalized Weierstrass function (GWF), that if said function has a fractal
dimension $\mathit{D}$ then the $\beta $-fractional derivative yields a new
function of fractal dimension $\mathit{D}+\beta $. In a similar way the 
$\beta $-fractional integral of the generalized Weierstrass function yields a
new function of fractal dimension $\mathit{D}-\beta $. Using these
properties we can write the fractional differential equations of motion for
the complex process of interest.

\section{Generalized Weierstrass Function}

The Weierstrass function was the first exemplar of a function that is
continuous and non-differentiable. A number of generalizations of this
fractal function have been discussed in the literature
\cite{mandelbrot,West3,falconer}, the most general including random behavior. 
We refer to the latter as the generalized Weierstrass function (GWF) defined 
by Berry and Lewis\cite{Berry and Lewis} as the series: 
\begin{equation}
W(t)=\sum_{n=-\infty }^{\infty }\frac{(1-e^{i\gamma ^{n}t})e^{i\phi _{n}}}{\gamma ^{(2-D)n}}  \label{1}
\end{equation}
where $1<D<2$, $\gamma >1$ and $\phi _{n}$ is an arbitrary phase. Berry and
Lewis show using a combination of numerics and analysis that for these
values of the parameters the GWF is continuous but not differentiable, has
no characteristic scale, and almost certainly has a fractal dimension $D$
\cite{footnote}. This argument was given mathematical rigor by Mauldin and
Williams\cite{mauldin} somewhat later. Further, if the set of phases 
$\left\{ \phi _{n}\right\} $ are uniformly distributed on the interval 
$\left( 0,2\pi \right) $ then $W(t)$ is a stochastic fractal function. The
properties of a modified version of (\ref{1}) was considered by Falconer
\cite{falconer}.

It is well known that (\ref{1}) has a fractal graph, that is, the trail left
by the GWF in $\left( X,t\right) $-space is a fractal curve. The fractal
nature of such graphs are manifest in the power-law behavior of the
correlation between measurements separated by a time interval $\tau$
\cite{falconer}. Before we consider the increments in the GWF, that is, the
difference in measurements separated by a time $\tau $, let us examine the
scaling properties of (\ref{1}) 
\begin{equation}
W\left( \gamma t\right) =\gamma ^{2-D}W\left( t\right)  \label{a}
\end{equation}
which is obtained by relabeling the series index. The most general solution
to the scaling equation (\ref{a}) is\cite{West3,meakin}

\begin{equation}
W\left( t\right) =A\left( t\right) t^{\alpha }  \label{b}
\end{equation}
where $A\left( t\right) $ is a function periodic in the logarithm of $t$,
with period ln$\gamma $, and the power-law index is given by

\begin{equation}
\alpha =2-D.  \label{c}
\end{equation}
Thus, we see that the dominant behavior of the GWF is a power-law growth in
time, and the power-law index is determined by the fractional dimension, $D$.

It is possible to connect the expression for the fractional dimension, 
(\ref{c}), with the slope of the correlation function of the increments in the
GWF. Let us define such a correlation function in the following way 
\begin{equation}
C(\tau )=\langle |\Delta W(t,\tau )|^{2}\rangle _{\phi }  \label{4}
\end{equation}
where the increments of the GWF are defined by 
\begin{equation}
\Delta W(t,\tau )=W(t+\tau )-W(t)  \label{5}
\end{equation}
and the average, indicated by the brackets with a $\phi $ subscript in 
(\ref{4}) is taken over an ensemble of realizations of the phases $\left\{ 
\phi\right\}$ uniformly distributed on the interval $\left( 0,2\pi \right) $.
We refer to the function (\ref{5}) as the incremental GWF or IGWF. Inserting
the IGWF 
\begin{equation}
\Delta W\left( t,\tau \right) =\sum_{n=-\infty }^{\infty }\gamma ^{-\left(
2-D\right) n}(1-e^{i\gamma ^{n}\tau })e^{i\left( \gamma ^{n}t+\phi
_{n}\right) }  \label{6}
\end{equation}
into the phase average in (\ref{4}) and carrying out the phase average 
\begin{equation}
\left\langle \left( ...\right) \right\rangle _{\phi }=\prod_{n=-\infty
}^{\infty }\int_{0}^{2\pi }\frac{d\phi _{n}}{2\pi }\left( ...\right)
\label{7}
\end{equation}
which for a uniform distribution of phases yields

\begin{equation}
\left\langle e^{i\left( \phi _{n}-\phi _{n^{\prime }}\right) }\right\rangle
_{\phi }=\delta _{n,n^{\prime }},  \label{delta}
\end{equation}
the correlation function becomes

\begin{equation}
C\left( \tau \right) =2\sum_{n=-\infty }^{\infty }\gamma ^{-2\left(
2-D\right) n}\left( 1-\cos \gamma ^{n}\tau \right) .  \label{corr}
\end{equation}
Note, since the correlation of the IGWF are independent of the initial time 
$t$, these increments are a realization of a stationary stochastic process.

The dominant behavior of the IGWF correlation function (\ref{corr}) is
determined by the solution to the exact scaling relation

\begin{equation}
C\left( \gamma \tau \right) =\gamma ^{2\left( 2-D\right) }C\left( \tau
\right)  \label{scale}
\end{equation}
obtained from (\ref{corr}) by again relabeling the series. The most general
solution to (\ref{scale}) is 
\begin{equation}
C(\tau )=A(\tau )\tau ^{2\alpha }  \label{8}
\end{equation}
where as before $A(\tau )$ is a function periodic in the logarithm of its
argument and the power-law index is given by (\ref{c})\cite{West3,meakin}.
Note that $\alpha $ plays the same role here as the Hurst exponent, $H$,
does in random walk processes. In fact if $\alpha =H=1/2$, then the
correlation function increases linearly with time, so that the IGWF, $\Delta
W,$ would be a normal diffusive process with Gaussian statistics that is
stationary in time.

\section{Fractional Calculus and the IGWF}

We are interested in both the fractional integral and the fractional
derivative of the generalized Weierstrass function. However, in applying the
fractional calculus, we take note of the fact that the GWF is not a
stationary stochastic function. It is the IGWF that is stationary. Therefore
we apply the fractional calculus to the stationary IGWF and thereby avoid
some technical difficulties.

\subsection{Fractional integral of the IGWF}

Let us introduce the Riemann-Liouville definition of a fractional integral
of order $\beta $ of the GWF: 
\begin{equation}
{\cal D}^{(-\beta )}W\left( t\right) \equiv \frac{1}{\Gamma (\beta )}
\int_{-\infty }^{t}\frac{W(t^{\prime })dt^{\prime }}{(t-t^{\prime
})^{1-\beta }},  \label{10}
\end{equation}
where $0<\beta <1$. We have
for the fractional integral of the IGWF: 
\begin{equation}
\Delta W^{(-\beta )}(t,\tau )\equiv \frac{1}{\Gamma (\beta )}\int_{-\infty
}^{t}\frac{dt^{\prime }}{(t-t^{\prime })^{1-\beta }}\Delta W\left( t^{\prime
},\tau \right) ,  \label{12}
\end{equation}
so that inserting (\ref{6}) into (\ref{12}) yields

\begin{eqnarray}
\Delta W^{(-\beta )}(t,\tau ) &\equiv &\frac{1}{\Gamma (\beta )}
\sum_{n=-\infty }^{\infty }\frac{e^{i\phi _{n}}}{\gamma ^{\left( 2-D\right)
n}}(1-e^{i\gamma ^{n}\tau })  \nonumber \\
&&\times \left\{ \int_{-\infty }^{t}\frac{dt^{\prime }}{(t-t^{\prime
})^{1-\beta }}e^{i\gamma ^{n}t^{\prime }}\right\}  \label{incre}
\end{eqnarray}
The integral between the curly braces yields

\begin{equation}
\Gamma \left( \beta \right) \frac{e^{i\gamma ^{n}t}}{\gamma ^{\beta n}}
e^{-i\pi \beta /2}  \label{int}
\end{equation}
and therefore

\begin{equation}
\Delta W^{(-\beta )}(t,\tau )\equiv e^{-i\pi \beta /2}\sum_{n=-\infty
}^{\infty }\frac{e^{i\left( \phi _{n}+\gamma ^{n}t\right) }}{\gamma ^{\left(
2-D+\beta \right) n}}(1-e^{i\gamma ^{n}\tau }).  \label{incre2}
\end{equation}
Thus, we see that the fractional integral of the IGWF has the same form as
the original IGWF. The difference between (\ref{incre2}) and (\ref{6}), up
to an overall phase, is that $D\rightarrow D-\beta $. What is the proper
interpretation of this shifting of the parameter value?

To answer this question, we need to go back to the GWF. We shall address
this issue in the next section. For the time being, we limit ourselves to
noticing that the correlation function related to (\ref{incre2}) is given by

\begin{equation}
C\left( \tau \right) =2\sum_{n=-\infty }^{\infty }\gamma ^{-2\left(
2-D+\beta \right) n}\left( 1-\cos \gamma ^{n}\tau \right)  \label{corr2}
\end{equation}
and therefore, if (\ref{incre2}) corresponds to some difference of properly
defined GWFs, by virtue of the scaling index (\ref{c}), these new GWFs would
have a new fractional dimension given by $D^{\prime }=D-\beta $. In the next
section we shall make this argument rigorous.

\subsection{Fractional derivative of the IGWF}

The calculations in the case of the fractional derivative of the IGWF are
similar to those carried out in the case of the fractional integral of the
IGWF. Let us consider the Riemann-Liouville definition of the $\beta$
fractional derivative of the GWF: 
\begin{equation}
{\cal D}^{(\beta )}W(t)\equiv \frac{1}{\Gamma (1-\beta )}\frac{d}{dt}
\int_{-\infty }^{t}\frac{W(t^{\prime })dt^{\prime }}{(t-t^{\prime })^{\beta }}. \label{22}
\end{equation}
where again $0<\beta <1$. The fractional derivative of the IGWF result in 
\begin{equation}
\Delta W^{(\beta )}(t,\tau )=\frac{1}{\Gamma (1-\beta )}\frac{d}{dt}
\int_{-\infty }^{t}\frac{dt^{\prime }}{(t-t^{\prime })^{\beta }}\Delta
W^{(\beta )}(t^{\prime },\tau ).  \label{23}
\end{equation}
The expression (\ref{incre}) is now replaced with
\begin{eqnarray}
\Delta W^{(\beta )}(t,\tau ) &=&\frac{1}{\Gamma (\beta )}\sum_{n=-\infty
}^{\infty }\frac{e^{i\phi _{n}}}{\gamma ^{\left( 2-D\right) n}}(1-e^{i\gamma
^{n}\tau })  \nonumber \\
&&\times \left\{ \frac{d}{dt}\int_{-\infty }^{t}\frac{dt^{\prime }}
{(t-t^{\prime })^{1-\beta }}e^{i\gamma ^{n}t^{\prime }}\right\}
\label{incre3}
\end{eqnarray}
where the time derivative of the integral in curly braces is

\begin{equation}
i\Gamma \left( 1-\beta \right) \gamma ^{\beta n}e^{i\gamma ^{n}t}e^{-i\pi
\left( \beta -1\right) /2}  \label{int2}
\end{equation}
and therefore

\begin{equation}
\Delta W^{(\beta )}(t,\tau )\equiv e^{i\pi \beta /2}\sum_{n=-\infty
}^{\infty }\frac{e^{i\left( \phi _{n}+\gamma^{n}t\right) }}{\gamma ^{\left(
2-D-\beta \right) n}}(1-e^{i\gamma ^{n}\tau }).  \label{deriv}
\end{equation}
Thus, we see that the fractional derivative of the IGWF has the same form as
the original IGWF. The difference between (\ref{deriv}) and (\ref{6}), up to
an overall phase, is that $D\rightarrow D+\beta $.

Again after observing that the correlation function related to (\ref{deriv})
is

\begin{equation}
C\left( \tau \right) =2\sum_{n=-\infty }^{\infty }\gamma ^{-2\left(
2-D-\beta \right) n}\left( 1-\cos \gamma ^{n}\tau \right)  \label{corr3}
\end{equation}
we make the plausible conjecture that $D+\beta $ corresponds to the new
fractional dimension, $D^{\prime \prime }$, of some properly defined GWF.
Again this is made rigorous in the next section.

\section{What about the GWF?}

We now want to use the fractional integral and fractional derivative of the
IGWF to determine these same operations of the GWF. To accomplish this we
assume that the IGWF is a given, as are its fractional integral and
fractional derivative, but the GWF remains to be determined. Using the RHS
of (\ref{5}) and (\ref{6}) we obtain

\begin{eqnarray}
W\left( t+\tau \right) -W\left( t\right) &=&\sum_{n=-\infty }^{\infty }
\frac{e^{i\left( \gamma ^{n}t+\phi _{n}\right) }}{\gamma ^{\left( 2-D\right) n}} \nonumber \\
&&-\sum_{n=-\infty }^{\infty }\frac{e^{i\left( \gamma ^{n}\left( t+\tau
\right) +\phi _{n}\right) }}{\gamma ^{\left( 2-D\right) n}}  \label{diff}
\end{eqnarray}
from which we can define the function

\begin{equation}
f\left( t\right) =W\left( t\right) +\sum_{n=-\infty }^{\infty }\frac{
e^{i\left( \gamma ^{n}t+\phi _{n}\right) }}{\gamma ^{\left( 2-D\right) n}}
\label{funct}
\end{equation}
where $W\left( t\right) $ is assumed to be unknown, and (\ref{diff}) is
replaced with the condition

\begin{equation}
f\left( t+\tau \right) =f\left( t\right) .  \label{perod}
\end{equation}
The series in (\ref{funct}) is divergent, so that in order to regularize the
function we write

\begin{equation}
f_{N}\left( t\right) =W_{N}\left( t\right) +\sum_{n=-N}^{N}\frac{e^{i\left(
\gamma ^{n}t+\phi _{n}\right) }}{\gamma ^{\left( 2-D\right) n}}
\label{regfunct}
\end{equation}
where

\begin{eqnarray}
f\left( t\right) &=&\lim_{N \rightarrow \infty}f_{N}\left(
t\right)  \nonumber \\
W\left( t\right) &=&\lim_{N \rightarrow \infty}W_{N}\left(
t\right) .  \label{limits}
\end{eqnarray}
The constraint (\ref{perod}) is now replaced with

\begin{equation}
\lim_{N \rightarrow \infty}f_{N}\left( t+\tau \right) =
\lim_{N \rightarrow \infty}f_{N}\left( t\right)
\label{constraint}
\end{equation}
indicating that the regularized function is either periodic with period 
$\tau $ or is a constant in the limit.

We now wish to establish that knowing the IGWF uniquely determines the
function $W\left( t\right) $, and this is the GWF. To accomplish this we use
the constraint given by (\ref{constraint}). We present the analysis for the
case $f_{N}\left( t\right) =A_{N},$ where $A_{N}$ is a constant, so that 
(\ref{regfunct}) can be written

\begin{equation}
W_{N}\left( t\right) =A_{N}-\sum_{n=-N}^{N}\frac{e^{i\left( \gamma^{n}t+\phi _{n}\right) }}{\gamma ^{\left( 2-D\right) n}}.  \label{newGWF}
\end{equation}
In order to determine the constant in (\ref{newGWF}) we impose an additional
constraint on the function. Since we want the function $W\left( t\right) $
to be a fractal we require that $W_{N}\left( t\right) $ satisfy the scaling
condition,

\begin{equation}
\lim_{N \rightarrow \infty}W_{N}\left( \gamma t\right) =
\lim_{N \rightarrow \infty}bW_{N}\left( t\right)
\label{GWFscale}
\end{equation}
where $b$ is a constant. Note that we exclude the possibility that the
function $f_{N}\left( t\right) $ is periodic, since this will not satisfy
this additional requirement that the function also scales. Imposing this
scaling constraint on (\ref{newGWF}) we have

\begin{equation}
W_{N}\left( \gamma t\right) =A_{N}-\gamma ^{\left( 2-D\right)
}\sum_{n=-N+1}^{N-1}\frac{e^{i\left( \gamma ^{n}t+\phi _{n}\right) }}{\gamma
^{\left( 2-D\right) n}}  \label{rescale}
\end{equation}
so if we choose $b=\gamma ^{2-D}$ we can satisfy (\ref{GWFscale}) in the 
$N\rightarrow \infty $ limit if we also choose

\begin{equation}
A_{N}=\sum_{n=-N}^{N}\frac{e^{i\theta _{n}}}{\gamma ^{\left( 2-D\right) n}},
\label{constant}
\end{equation}
where $\left\{ \theta _{n}\right\} $ is a set of arbitrary phases, because

\begin{equation}
\lim_{N \rightarrow \infty}A_{N}= \lim_{N \rightarrow \infty}\gamma ^{2-D}A_{N}.  \label{factor}
\end{equation}
Thus, the divergences that required the regularization exactly cancel in 
(\ref{rescale}) with the choice of constraint (\ref{constant}) and we obtain
the function

\begin{equation}
W\left( t\right) =\lim_{N \rightarrow \infty}W_{N}\left(
t\right) =\sum_{n=-\infty }^{\infty }\frac{\left( 1-e^{i\gamma ^{n}t}\right)
e^{i\phi _{n}}}{\gamma ^{\left( 2-D\right) n}}  \label{atlast}
\end{equation}
which is the GWF, where we have associated the set of phases with those used
earlier in the GWF.

This argument can also be applied to the fractional integral of the IGWF
given by (\ref{incre2}) so that we also have for the fractional integral of
the GWF

\begin{equation}
W^{(-\beta )}(t,\tau )\equiv e^{-i\pi \beta /2}\sum_{n=-\infty }^{\infty }
\frac{e^{i\phi _{n}}}{\gamma ^{\left( 2-D+\beta \right) n}}(1-e^{i\gamma
^{n}t}).  \label{GWFint}
\end{equation}
In the same way the fractional derivative of the IGWF given by (\ref{deriv})
implies that the fractional derivative of the GWF is

\begin{equation}
W^{(\beta )}(t,\tau )\equiv e^{i\pi \beta /2}\sum_{n=-\infty }^{\infty }
\frac{e^{i\phi _{n}}}{\gamma ^{\left( 2-D-\beta \right) n}}(1-e^{i\gamma
^{n}t}).  \label{GWFder}
\end{equation}
Thus, our earlier remarks regarding the fractional calculus applied to the
IGWF apply equally well to the GWF.

\section{Conclusions}

We have determined the fractional dimension of both fractional integral and
derivative of the GWF. Our result reads in these two cases: 
\begin{eqnarray}
&&{\rm Dim}[{\cal D}^{(-\beta )}W] = {\rm Dim}[W]-\beta ,  \label{dim1} \\
&&{\rm Dim}[{\cal D}^{(\beta )}W] = {\rm Dim}[W]+\beta .  \label{dim2}
\end{eqnarray}
This result can be easily interpreted noticing that the fractional dimension
gives information about the degree of irregularity of the function under
analysis. We have demonstrated that carrying out a fractional integral of
the GWF means decreasing its fractional dimension and therefore smoothing
the process, while carrying out a fractional derivative means increasing the
fractional dimension and therefore making the process and its increments
more irregular.

A related result was obtained by Kolwankar and Gangal\cite{kol1,kol2}, but
required the introduction of a local fractional derivative (LFD), that is, a
fractional derivative defined such that its non-local character is removed.
They find that the LFD of a Weierstrass function, that is, the imaginary
part of (\ref{1}) with all $\phi _{n}=0$, exists up to ''critical order'' 
$2-D$ and not so for orders between $2-D$ and $1$, where $D$ $(1<D<2)$ is the
box counting dimension of the graph of the function. It is possible to show
that our result is in complete agreement with that of Kolwankar and Gangal
\cite{kol1,kol2}. To this aim, let us consider the inequality

\begin{equation}
D_{T}<D<D_{E}  \label{ineq}
\end{equation}
where $D_{T}$ is the topological dimension and $D_{E}$ is the embedding
dimension\cite{mandelbrot}. The condition (\ref{ineq}) needs to be fulfilled
by any function in order to be a fractal. In the specific case of the graph
of a fractal function, $D_{T}=1$ and $D_{E}=2$. The same condition (\ref
{ineq}) must also hold true for the fractional derivative (fractional
integral) of the function, in this case the GWF, in order to preserve its
fractal properties. Therefore, for the fractional derivative of GWF, we have:

\begin{equation}
D^{\prime \prime }<2\Rightarrow D+\beta <2\Rightarrow \beta <2-D.
\label{dimen}
\end{equation}
This means that the generalized Weierstrass function is fractionally
differentiable for all orders less than $2-D$, which is the result of
Kolwankar and Gangal\cite{kol1,kol2}.

Moreover, the same kind of reasoning can be applied to the fractional
integral of the GWF. In this case, the meaningful condition reads:

\begin{equation}
D^{\prime }>1\Rightarrow D-\beta >1\Rightarrow \beta <D-1.  \label{dimen2}
\end{equation}
This means that the generalized Weierstrass function is fractionally
integrable for all orders less than $D-1.$

Finally, we go back to the main issue, that being, a possible equation of
evolution for a complex system, exhibiting fractal behavior and perhaps 
representable by the increments of a generalized Weierstrass function.

Consider the function

\begin{equation}
f\left( t\right) =\sum_{n=-\infty }^{\infty }A_{n}\left( 1-e^{i\omega_{n}t}\right) e^{i\phi _{n}}  \label{stochas}
\end{equation}
where $\left\{ \omega _{n}\right\} $ is a set of frequencies, $\left\{ \phi
_{n}\right\} $ is a set of random phase confined to the interval $\left(
0,2\pi \right) $, and $A_{n}$ is a real amplitude. Thus, $f\left( t\right)$
is a stochastic function of time with the definite initial condition 
$f\left( 0\right) =0$. If we now use (\ref{stochas}) as a driving force for a
fractional stochastic equation we can write

\begin{equation}
{\cal D}^{(\alpha )}F(t)=f\left( t\right)  \label{fracdiff}
\end{equation}
as our equation of motion. The solution to (\ref{fracdiff}) is formally
given by the inverse equation

\begin{equation}
F(t)={\cal D}^{(-\alpha )}f\left( t\right)  \label{fracint}
\end{equation}
so that using the definition of the fractional integral, (\ref{10}), we
obtain the explicit form of the solution

\begin{equation}
F\left( t\right) =\sum_{n=-\infty }^{\infty }\frac{A_{n}}{\omega_{n}^{\alpha }}\left( 1-e^{i\omega _{n}t}\right) e^{i\phi _{n}^{\prime }}
\label{fracsol}
\end{equation}
where we have absorbed the overall phase into the new random phases $\{
\phi_{n}^{\prime }\}$. Therefore, if we choose the parameter
values $\omega _{n}=\gamma ^{n}$, $A_{n}=1$ and $\alpha =2-D$, the solution
to the fractional stochastic differential equation, (\ref{fracdiff}), is the
GWF.

In addition if we choose the coefficients in the series representation of
the random driving force by $A_{n} \propto 1/\omega _{n}$ then 
(\ref{stochas}) is a realization of a complex Brownian motion process, which is
to say that the statistics are two dimensional Gaussian and the spectrum is
an inverse square frequency. In this case if we also choose for the
parameters in the solution to the fractional stochastic differential
equation $\omega _{n}=\gamma ^{n}$ and $\alpha = 1 - D$, then
the solution is again a GWF. In addition, as pointed out by Mandelbrot\cite
{mandelbrot} (page 390), the GWF with random coefficients is a good
approximation to a fractional Brownian function.

\end{multicols}


\begin{thebibliography}{99}
\bibitem{mandelbrot}  B.B. Mandelbrot, \textit{The Fractal Geometry of Nature}, W.H. Freeman and Co., San Francisco (1977).

\bibitem{West3}  B.J. West, \textit{Physiology, Promiscuity and Prophecy at
the Millennium: A Tale of Tails}, Studies of Nonlinear Phenomena in the
Life Sciences Vol. 9, World Scientific, Singapore (1998).

\bibitem{meakin}  P. Meakin, \textit{Fractals, scaling and growth far from
equilibrium}, Cambridge Nonlinear Science Series 5, Cambridge University
Press, Cambridge (1998).

\bibitem{schroeder}  M. Schroeder, \textit{Fractals, Chaos, Power Laws},
W.H. Freeman and Comp., New York (1991).

\bibitem{richardson}  L.F. Richardson, Proc. Roy. Soc. London A 110 
(1926) 709.

\bibitem{shlesinger}  M.F. Shlesinger, B.J. West and J. Klafter, Phys. Rev.
Lett. 58 (1987) 1100.

\bibitem{levy}  P. L\'{e}vy, \textit{Calcul des probabilities},
Guthier-Villars, Paris (1925); \textit{Th\'{e}orie de l'addition des
variables al\'{e}atoires}, Guthier-Paris (1937).

\bibitem{hughes}  B.D. Hughes,\textit{\ Random Walks and Random
Environments, Volume 1: Random Walks}, Oxford Science Publications,
Clarendon Press, Oxford (1995).

\bibitem{montroll1}  E.W. Montroll and B.J. West, in \textit{Fluctuation Phenomena}, pp. 61-206, Editors E.W. Montroll and J.L. Lebowitz, Second Edition,
North-Holland Personal Library, North-Holland, Amsterdam (1987)

\bibitem{montroll2}  E.W. Montroll and M.F. Shlesinger, in 
\textit{Nonequilibrium Phenomena II: From
Stochastics to Hydrodynamics}, pp. 1-121, Editors E.W. Montroll and J.L.
Lebowitz, North-Holland, Amsterdam (1983).

\bibitem{langevin}  P. Langevin, C.R. Acad. Sci. 530, Paris (1908).

\bibitem{perrin}  J. Perrin, ''Mouvement brownien et r\'{e}alit\'{e} mol\'{e}
culaire'', Annales de chimie et de physique VIII 18, 5-114: Translated by F.
Soddy as \textit{Brownian Movement and Molecular Reality}, Taylor and
Francis, London.

\bibitem{doob}  J.L. Doob, Ann. Math. 43 (1942) 351.

\bibitem{orn}  G.E. Uhlenbeck and L.S. Ornstein, Phys. Rev. 36 (1930) 823.

\bibitem{grigolini}  P. Grigolini, A. Rocco and B.J. West, to appear in Phys.
Rev. E.

\bibitem{West1}  P. Allegrini, P. Grigolini and B.J. West, Phys. Rev. E 54 (1996) 4760.

\bibitem{compte}  A. Compte, Phys. Rev. E 53 (1996) 4191.

\bibitem{chaves}  A.S. Chaves, Phys. Lett. A 239 (1998) 13.

\bibitem{hosking}  J.T.M. Hosking, Biometrika 68 (1981) 165.

\bibitem{bernan}  J. Bernan, \textit{Statistics of Long-Memory Processes},
Monograph on Statistics and Applied Probability 61, Chapman and
Hall, New York (1994).

\bibitem{West2}  B.J. West and P. Grigolini, in \textit{Applications of Fractional
Calculus in Physics}, Editor R. Hilfer, World Scientific, Singapore (1998).

\bibitem{Nonn}  W.G. Gl\"{o}ckle and T.F. Nonnenmacher, Rheol. Acta 33 (1994) 337; J. Stat. Phys. 71 (1993) 741.

\bibitem{Berry}  M. Berry, J. Phys. A: Math. Gen. 12 (1979) 781.

\bibitem{Wyss}  W.R. Schneider and W. Wyss, J. Math. Phys. 30 (1989) 134.

\bibitem{footnote2}  This view is not free of criticism. Many attempts have
been made to reproduce the thermodynamical properties of the Universe from
Newton's laws, but it seems that this can not be done in a satisfactory way
unless non-dynamical assumptions like coarse graining, the Markov
approximation or Van Hove approaches are made \cite{grigolini}.

\bibitem{monin}  A.S. Monin and A.M. Yaglom, \textit{Statistical Fluid
Mechanics: Mechanics of Turbulence, Volumes 1 and 2}, MIT Press, Cambridge,
Mass. (1971).

\bibitem{falconer}  K. Falconer, \textit{Fractal Geometry}, Wiley, New York
(1990).

\bibitem{Berry and Lewis}  M.V. Berry and Z.V. Lewis, Proc. Roy. Soc. Lond. A 370 (1980) 459.

\bibitem{footnote}  In the following we use the term fractional dimension to
avoid specious arguments over whether $D$ is the box counting dimension, the
Hausdorff-Besicovitch dimension or the fractal dimension. For our immediate
purposes it is sufficient that $D$ is not an integer and is in the interval 
$1<D<2$.

\bibitem{mauldin}  R.D. Mauldin and S.C. Williams, Trans. Am. Math. Soc. 298 (1986) 793.

\bibitem{kol1}  K.M. Kolwankar and A.D. Gangal, Chaos 6 (1996) 505.

\bibitem{kol2}  K.M. Kolwankar and A.D. Gangal, Pramana J. Phys. 48 (1997) 49.

\end{thebibliography}
\end{document}